\documentclass[a4paper,fleqn]{cas-sc}

\usepackage[numbers]{natbib}
\usepackage[]{lineno}

\usepackage{lipsum}
\usepackage{amsmath}
\usepackage{float}

\usepackage{textcomp}

\usepackage[separate-uncertainty, multi-part-units=single]{siunitx} 

\def\tsc#1{\csdef{#1}{\textsc{\lowercase{#1}}\xspace}}
\tsc{WGM}
\tsc{QE}

\newcommand{\ladder}[1]{\ensuremath{\mathrm{L_{#1}}}}
\newcommand{\Plotsize}{0.55} 
\newcommand{\PlotsizeL}{0.7} 

\begin{document}
\let\WriteBookmarks\relax
\def\floatpagepagefraction{1}
\def\textpagefraction{.001}
\newcommand{\dndeta}{\mbox{d$N_{\rm ch}$/d$\eta$}\xspace}

\newcommand{\sphenix}{\mbox{sPHENIX}\xspace}
\newcommand{\phenix}{\mbox{PHENIX}\xspace}
\newcommand{\phobos}{\mbox{PHOBOS}\xspace}
\newcommand{\brahms}{\mbox{BRAHMS}\xspace}
\newcommand{\alice}{\mbox{ALICE}\xspace}
\newcommand{\cms}{\mbox{CMS}\xspace}
\newcommand{\atlas}{\mbox{ATLAS}\xspace}
\newcommand{\starexp}{\mbox{STAR}\xspace}

\newcommand{\mvtx}{\mbox{MVTX}\xspace}
\newcommand{\intt}{\mbox{INTT}\xspace}
\newcommand{\tpc}{\mbox{TPC}\xspace}
\newcommand{\tpot}{\mbox{TPOT}\xspace}
\newcommand{\emcal}{\mbox{EMCAL}\xspace}
\newcommand{\ihcal}{\mbox{iHCAL}\xspace}
\newcommand{\ohcal}{\mbox{oHCAL}\xspace}
\newcommand{\mbd}{\mbox{MBD}\xspace}
\newcommand{\zdc}{\mbox{ZDC}\xspace}
\newcommand{\sepd}{\mbox{sEPD}\xspace}
\newcommand{\smd}{\mbox{SMD}\xspace}

\newcommand{\minbias}{\textsc{Min. Bias}\xspace}

\newcommand{\hic}{\mbox{$A$$+$$A$}\xspace}
\newcommand{\ee}{\mbox{$e^{+}$$+$$e^{-}$}\xspace}
\newcommand{\ep}{\mbox{$e$$+$$p$}\xspace}
\newcommand{\eA}{\mbox{$e$$+$$A$}\xspace}
\newcommand{\AuAu}{\mbox{Au$+$Au}\xspace}
\newcommand{\uu}{\mbox{U$+$U}\xspace}
\newcommand{\apa}{\mbox{A$+$A}\xspace}
\newcommand{\auau}{\mbox{Au$+$Au}\xspace} 
\newcommand{\alal}{\mbox{Al$+$Al}\xspace} 
\newcommand{\oo}{\mbox{O$+$O}\xspace} 
\newcommand{\agag}{\mbox{Ag$+$Ag}\xspace} 
\newcommand{\cucu}{\mbox{Cu$+$Cu}\xspace} 
\newcommand{\xexe}{\mbox{Xe$+$Xe}\xspace} 
\newcommand{\raa}{\mbox{$R_{AA}$}\xspace}
\newcommand{\pbpb}{\mbox{Pb$+$Pb}\xspace} 
\newcommand{\ppb}{\mbox{p$+$Pb}\xspace} 
\newcommand{\pdau}{\mbox{$p(d)$$+$Au}\xspace} 
\newcommand{\aj}{\mbox{$A_J$}\xspace} 
\newcommand {\pp}{\mbox{$p$$+$$p$}\xspace}
\newcommand {\ppbar}{\mbox{$p$$+$$\overline{p}$}\xspace}
\newcommand{\pT}{\mbox{${p_T}$}\xspace}
\newcommand{\jpsi}{\mbox{$J/\psi$}\xspace}
\newcommand{\sqrts}{\mbox{$\sqrt{s}$}\xspace}
\newcommand{\sqrtsnn}{\mbox{$\sqrt{s_{\scriptscriptstyle NN}}$}\xspace}
\newcommand{\npart}{$N_{\mathrm{part}}$\xspace}
\newcommand{\ncoll}{$N_{\mathrm{coll}}$\xspace}
\newcommand{\qgp}{\mbox{quark-gluon plasma}\xspace}
\newcommand{\jt}{\mbox{$J_T$}} \newcommand{\qhat}{\mbox{$\hat{q}$}\xspace}
\newcommand{\Qsqr}{\mbox{$Q^2$}} \newcommand{\CuCu}{\mbox{Cu$+$Cu}\xspace}
\newcommand{\PbPb}{\mbox{Pb$+$Pb}\xspace} 
\newcommand{\pPb}{\mbox{$p$$+$Pb}\xspace}
\newcommand{\gjet}{\mbox{$\gamma$-jet}\xspace}
\newcommand{\Qmax}{\mbox{$Q_{\max}$}} \newcommand{\ET}{\mbox{$E_T$}}
\newcommand{\Et}{\mbox{$E_T$}} \newcommand{\kt}{\mbox{$k_T$}}
\newcommand{\RAA}{\mbox{$R_{AA}$}\xspace} \newcommand{\IAA}{\mbox{$I_{AA}$}}
\newcommand{\pt}{\mbox{${p_{\mathrm{T}}}$}\xspace}
\newcommand{\nb}{\mbox{nb$^{-1}$}\xspace}
\newcommand{\pb}{\mbox{pb$^{-1}$}\xspace}
\newcommand{\fb}{\mbox{fb$^{-1}$}\xspace}

\newcommand{\Dzero}{$D^{0}$\xspace}
\newcommand{\dTokpi}{$D^{0}\rightarrow K^-\pi^+$\xspace}
\newcommand{\mkpi}{$m_{K^-\pi^+}$\xspace}
\newcommand {\bbbar}{\mbox{$b\overline{b}$}\xspace}
\newcommand {\ccbar}{\mbox{$c\overline{c}$}\xspace}

\newcommand{\highpt}{high-${\rm p_{_{T}}}$}
\newcommand{\lessim}{{\stackrel{<}{\sim}}} \newcommand{\eqnpt}{p_T}
\newcommand{\dAu}{\mbox{$d$$+$Au}\xspace}
\newcommand{\pAu}{\mbox{$p$$+$Au}\xspace}
\newcommand{\pau}{\mbox{$p$$+$Au}\xspace}
\newcommand{\pAl}{\mbox{$p$$+$Al}\xspace}
\newcommand{\gevsq}{\mbox{${\rm~GeV}^2$}\xspace}
\newcommand{\fastjet}{\mbox{\sc FastJet}\xspace}
\newcommand{\hepmc}{\mbox{\sc HepMC2}\xspace}
\newcommand{\geant}{\mbox{\sc Geant4}\xspace}
\newcommand{\antikt}{\mbox{anti-$k_T$}\xspace}
\newcommand{\pythia}{\mbox{\sc Pythia8}\xspace}
\newcommand{\funforall}{\mbox{\sc Fun4All}\xspace}
\newcommand{\kfparticle}{\mbox{\sc KFParticle}\xspace}
\newcommand{\decayfinder}{\mbox{\sc DecayFinder}\xspace}
\newcommand{\hftrackeff}{\mbox{\sc HFTrackEfficiency}\xspace}
\newcommand{\rapgap}{\mbox{\sc Rapgap}\xspace}
\newcommand{\milou}{\mbox{\sc Milou}\xspace}
\newcommand{\pyquen}{\mbox{\tt Pyquen}\xspace}
\newcommand{\hijing}{\mbox{\tt HIJING}\xspace}
\newcommand{\ampt}{\mbox{\tt AMPT}\xspace}
\newcommand{\epos}{\mbox{\tt EPOS4}\xspace}
\newcommand{\jewel}{\mbox{\tt Jewel}\xspace}
\newcommand{\roofit}{\mbox{\sc RooFit}\xspace}
\newcommand{\roounfold}{\mbox{\sc RooUnfold}\xspace}
\newcommand{\beetle}{\mbox{\sc Beetle}\xspace}
\newcommand{\gj}{\mbox{$\gamma$+jet}\xspace}
\newcommand{\gh}{\mbox{$\gamma$+hadron}\xspace}
\newcommand{\martinimusic}{\mbox{\sc Martini+Music}\xspace}
\newcommand{\martini}{\mbox{\sc Martini}\xspace}
\newcommand{\music}{\mbox{\sc Music}\xspace}
\newcommand{\Ephenix}{Electron-Ion Collider (EIC) detector built
  around the BaBar magnet and sPHENIX calorimetry\xspace}
\newcommand{\ephenix}{EIC detector built around the BaBar magnet and
  sPHENIX calorimetry\xspace} 
\newcommand{\refdesign}{reference design\xspace}
\newcommand{\refconfig}{reference configuration\xspace}
\newcommand{\dijet}{\mbox{dijet}\xspace}
\newcommand{\fake}{\mbox{fake}\xspace}
\newcommand{\fast}{\mbox{fast}\xspace}
\newcommand{\veryfast}{\mbox{very fast}\xspace}
\newcommand{\epem}{\mbox{$e^+e^-$}\xspace}
\newcommand{\onewidth}{0.6\linewidth}
\newcommand{\twowidth}{0.48\linewidth}
\newcommand{\threewidth}{0.32\linewidth}

\newcommand{\egoing}{\mbox{electron-going}\xspace}
\newcommand{\hgoing}{\mbox{hadron-going}\xspace}
\newcommand{\egodir}{electron-going direction\xspace}
\newcommand{\hgodir}{hadron-going direction\xspace}
\newcommand{\bigcell}[2]{\begin{tabular}{@{}#1@{}}#2\end{tabular}}

\def\sPlot{\mbox{\em sPlot}\xspace}
\def\sPlots{\mbox{\em sPlots}\xspace}
\def\sWeights{\mbox{\em sWeights}\xspace}

\newcommand{\tev}{\ensuremath{\mathrm{\,Te\kern -0.1em V}}\xspace}
\newcommand{\gev}{\ensuremath{\mathrm{\,Ge\kern -0.1em V}}\xspace}
\newcommand{\mev}{\ensuremath{\mathrm{\,Me\kern -0.1em V}}\xspace}
\newcommand{\kev}{\ensuremath{\mathrm{\,ke\kern -0.1em V}}\xspace}
\newcommand{\ev}{\ensuremath{\mathrm{\,e\kern -0.1em V}}\xspace}
\newcommand{\gevc}{\ensuremath{{\mathrm{\,Ge\kern -0.1em V\!/}c}}\xspace}
\newcommand{\mevc}{\ensuremath{{\mathrm{\,Me\kern -0.1em V\!/}c}}\xspace}
\newcommand{\gevcc}{\ensuremath{{\mathrm{\,Ge\kern -0.1em V\!/}c^2}}\xspace}
\newcommand{\gevgevcccc}{\ensuremath{{\mathrm{\,Ge\kern -0.1em V^2\!/}c^4}}\xspace}
\newcommand{\mevcc}{\ensuremath{{\mathrm{\,Me\kern -0.1em V\!/}c^2}}\xspace}

\def\km   {\ensuremath{\rm \,km}\xspace}
\def\m    {\ensuremath{\rm \,m}\xspace}
\def\cm   {\ensuremath{\rm \,cm}\xspace}
\def\cma  {\ensuremath{{\rm \,cm}^2}\xspace}
\def\mm   {\ensuremath{\rm \,mm}\xspace}
\def\mma  {\ensuremath{{\rm \,mm}^2}\xspace}
\def\mum{\ensuremath{\mu\mathrm{m}}\xspace}
\def\muma {\ensuremath{\rm \,\mu\mathrm{m}^2}\xspace}
\def\nm   {\ensuremath{\rm \,nm}\xspace}
\def\fm   {\ensuremath{\rm \,fm}\xspace}
\def\barn{\ensuremath{\rm \,b}\xspace}
\def\barnhyph{\ensuremath{\rm -b}\xspace}
\def\mbarn{\ensuremath{\rm \,mb}\xspace}
\def\mub{\rm \,\textmu b\xspace}
\def\mbarnhyph{\ensuremath{\rm -mb}\xspace}
\def\nb {\ensuremath{\rm \,nb}\xspace}
\def\invnb {\ensuremath{\mbox{\,nb}^{-1}}\xspace}
\def\pb {\ensuremath{\rm \,pb}\xspace}
\def\invpb {\ensuremath{\mbox{\,pb}^{-1}}\xspace}
\def\fb   {\ensuremath{\mbox{\,fb}}\xspace}
\def\invfb   {\ensuremath{\mbox{\,fb}^{-1}}\xspace}
\def\khz   {\ensuremath{\mbox{\,kHz}}\xspace}
\def\mhz   {\ensuremath{\mbox{\,MHz}}\xspace}
\def\ghz   {\ensuremath{\mbox{\,GHz}}\xspace}
\def\mbs   {\ensuremath{\mbox{\,Mb/s}}\xspace}
\def\gbs   {\ensuremath{\mbox{\,Gb/s}}\xspace}
\def\tbs   {\ensuremath{\mbox{\,Tb/s}}\xspace}

\def\sec  {\ensuremath{\rm {\,s}}\xspace}
\def\ms   {\ensuremath{{\rm \,ms}}\xspace}
\def\mus  {\rm \,\textmu s\xspace}
\def\ns   {\ensuremath{{\rm \,ns}}\xspace}
\def\ps   {\ensuremath{{\rm \,ps}}\xspace}
\def\fs   {\ensuremath{\rm \,fs}\xspace}

\def\mhz  {\ensuremath{{\rm \,MHz}}\xspace}
\def\khz  {\ensuremath{{\rm \,kHz}}\xspace}
\def\hz   {\ensuremath{{\rm \,Hz}}\xspace}

\def\invps{\ensuremath{{\rm \,ps^{-1}}}\xspace}

\def\yr   {\ensuremath{\rm \,yr}\xspace}
\def\hr   {\ensuremath{\rm \,hr}\xspace}
\def\degc {\ensuremath{^\circ}{C}\xspace}
\def\degk {\ensuremath {\rm K}\xspace}

\def\Xrad {\ensuremath{X_0}\xspace}
\def\NIL{\ensuremath{\lambda_{int}}\xspace}
\def\mip {MIP\xspace}
\def\neutroneq {\ensuremath{\rm \,n_{eq}}\xspace}
\def\neqcmcm {\ensuremath{\rm \,n_{eq} / cm^2}\xspace}
\def\kRad {\ensuremath{\rm \,kRad}\xspace}
\def\MRad {\ensuremath{\rm \,MRad}\xspace}
\def\ci {\ensuremath{\rm \,Ci}\xspace}
\def\mci {\ensuremath{\rm \,mCi}\xspace}

\def\sx    {\ensuremath{\sigma_x}\xspace}    
\def\sy    {\ensuremath{\sigma_y}\xspace}   
\def\sz    {\ensuremath{\sigma_z}\xspace}    

\newcommand{\stat}{\ensuremath{\mathrm{(stat)}}\xspace}
\newcommand{\syst}{\ensuremath{\mathrm{(syst)}}\xspace}
\newcommand{\model}{\ensuremath{\mathrm{(model)}}\xspace}

\newcommand{\ten}[1]{\ensuremath{\times 10^{#1}}}
\def\order{{\ensuremath{\cal O}}\xspace}
\newcommand{\chisq}{\ensuremath{\chi^2}\xspace}
\newcommand{\erfc}[1]{\ensuremath{\rm{Erfc}(#1)}\xspace}

\def\deriv {\ensuremath{\mathrm{d}}}

\def\gsim{{~\raise.15em\hbox{$>$}\kern-.85em
          \lower.35em\hbox{$\sim$}~}\xspace}
\def\lsim{{~\raise.15em\hbox{$<$}\kern-.85em
          \lower.35em\hbox{$\sim$}~}\xspace}

\newcommand{\DR}{\ensuremath{\Delta \text{R}}}
\newcommand{\DPhi}{\ensuremath{\Delta\phi}}
\newcommand{\DEta}{\ensuremath{\Delta\eta}}

\shorttitle{Beam test results of the Intermediate Silicon Tracker for sPHENIX}

\shortauthors{<C. W. Shih, {\it et al.}>}  

\title[mode=title]{Beam test results of the Intermediate Silicon Tracker for sPHENIX}  



%

\author[NCU,RIKEN,RBRC]{C.~W.~Shih}
\author[RIKEN,RBRC]{G.~Nukazuka}[]\cormark[1]\ead{genki.nukazuka@riken.jp}
\author[NWU,RIKEN]{Y.~Sugiyama}[]
\author[RIKEN,RBRC]{Y.~Akiba}[]
\author[BNL]{D.~Cacace}[]
\author[RIKEN,RBRC]{H.~En'yo}[]
\author[NWU,RIKEN]{T.~Hachiya}[]
\author[JAEA,RIKEN]{S.~Hasegawa}[]
\author[NWU,RIKEN]{M.~Hata}[]
\author[Rikkyo, RIKEN]{H.~Imai}[]
\author[NCU]{C.~M.~Kuo}[]
\author[NWU,RIKEN]{M.~Morita}[]
\author[RIKEN,RBRC]{I.~Nakagawa}[]
\author[Rikkyo,RIKEN]{Y.~Nakamura}[]
\author[Rikkyo,RIKEN]{G.~Nakano}[]
\author[NWU,RIKEN]{Y.~Namimoto}[]
\author[BNL]{R.~Nouicer}[]
\author[BNL]{R.~Pisani}[]
\author[NWU,RIKEN]{M.~Shibata}[]
\author[NWU,RIKEN]{M.~Shimomura}[]
\author[NWU,RIKEN]{R.~Takahama}[]
\author[RARiS]{K.~Toho}[]
\author[NWU,RIKEN]{H.~Tsujibata}[]
\author[RARiS]{M.~Tsuruta}[]
\author[NWU,RIKEN]{M.~Watanabe}[]






\affiliation[NCU]{organization={Department of Physics and Center for High Energy and High Field Physics, National Central University},
            addressline={No.300, Zhongda Rd., Zhongli Dist.}, 
            city={Taoyuan City},
            postcode={32001}, 
            country={Taiwan}}

\affiliation[RIKEN]{organization={Nishina Center for Accelerator-Based Science, RIKEN},
            addressline={2-1 Hirosawa}, 
            city={Wako},
            postcode={351-0198}, 
            state={Saitama},
            country={Japan}}

\affiliation[RBRC]{organization={RIKEN BNL Research Center},
            addressline={20 Pennsylvania Avenue}, 
            city={Upton},
            postcode={11973}, 
            state={NY},
            country={U.S.A.}}       

\affiliation[NWU]{organization={Department of Mathematical and Physical Sciences, Nara Women's University},
            addressline={Kitauoya-Higashimachi}, 
            city={Nara},
            postcode={630-8506}, 
            state={Nara},
            country={Japan}}  

\affiliation[BNL]{organization={Physics Department, Brookhaven National Laboratory},
            addressline={20 Pennsylvania Avenue}, 
            city={Upton},
            postcode={11973}, 
            state={NY},
            country={U.S.A.}}

\affiliation[JAEA]{organization={Advanced Science Research Center, Japan Atomic Energy Agency},
            addressline={2-4 Shirakata Shirane}, 
            city={Tokai-mura, Naka-gun},
            postcode={319-1195}, 
            state={Ibaraki},
            country={Japan}}

\affiliation[Rikkyo]{organization={Department of Physics, Rikkyo University},
             addressline={3-34-1 Nishi-Ikebukuro, Toshima}, 
            city={Tokyo},
            postcode={171-8501}, 
            state={Tokyo},
            country={Japan}}

\affiliation[RARiS]{organization={Research Center for Accelerator and Radioisotope Science, Tohoku University},
             addressline={1-2-1 Mikamine, Taihaku}, 
            city={Sendai},
            postcode={982-0826}, 
            state={Miyagi},
            country={Japan}}





\cortext[1]{Corresponding author}



\begin{abstract}
The Intermediate Silicon Tracker (INTT), a two-layer barrel silicon strip tracker, is a key component of the tracking system for sPHENIX at the Relativistic Heavy Ion Collider (RHIC) at Brookhaven National Laboratory. The INTT is designed to enable the association of reconstructed tracks with individual RHIC bunch crossings. To evaluate the performance of preproduction INTT ladders and the readout chain, a beam test was conducted at the Research Center for Accelerator and Radioisotope Science, Tohoku University, Japan. This paper presents the performance of the INTT evaluated through studies of the signal-to-noise ratio, residual distribution, spatial resolution, hit-detection efficiency, and multiple track reconstruction.
\end{abstract}




\maketitle


\section{Introduction}

\subsection{The \sphenix experiment}

\sphenix~\cite{PHENIX:2015siv} is a new major detector at the Relativistic Heavy Ion Collider (RHIC)~\cite{HARRISON2003235} at Brookhaven National Laboratory (BNL). The primary goal of \sphenix is to study the nature of the strongly interacting quark-gluon plasma (QGP) and cold quantum chromodynamics (QCD), including proton spin physics, through high-precision and high-statistics measurements of hard-probe observables~\cite{Aprahamian:2015qub,Belmont:2023fau}. Designed as a state-of-the-art jet detector, \sphenix features barrel electromagnetic and hadronic calorimeters\textemdash available for the first time at mid-pseudorapidity at RHIC\textemdash thereby enabling fully reconstructed jet measurements.
In addition, the detector includes a precision tracking system comprising four subsystems, along with a refurbished 1.4 Tesla superconducting solenoid magnet from the \textit{BABAR} experiment~\cite{AUBERT20021}, supporting detailed studies of jet substructure and heavy-flavor observables.
Together, these components provide full azimuthal coverage and pseudorapidity acceptance of $|\eta|<1.1$ for collision vertices located within $\pm10$\cm of the nominal interaction point along the beam axis.

\subsection{The Intermediate Silicon Tracker (INTT)}
\label{sec_INTT_intro}
\begin{figure}[ht]
    \centering
    \includegraphics[width=\Plotsize\linewidth]{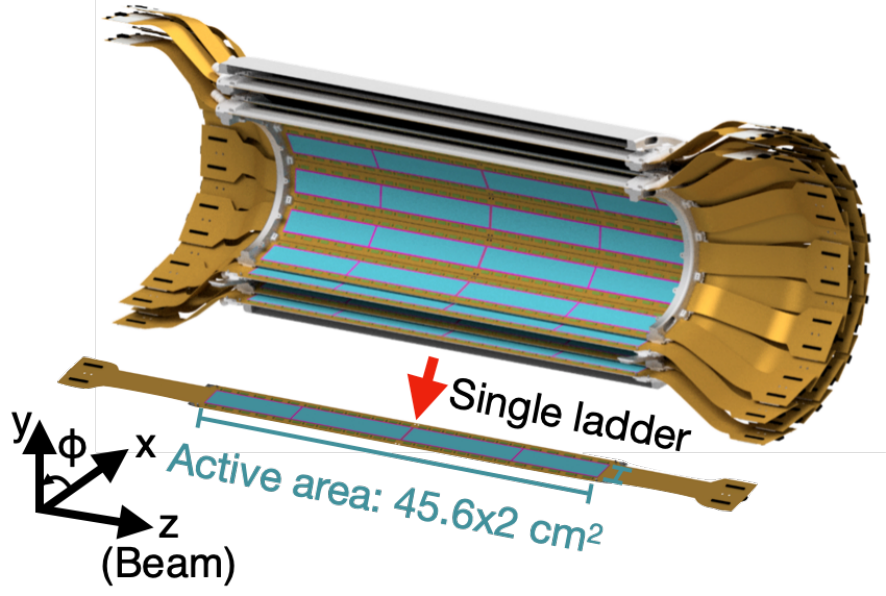}
    \caption{The schematic drawing of the INTT barrel and a ladder. The active areas of the ladders are shown as cyan boxes. The coordinate system used for the \sphenix experiment is also shown.}
    \label{fig:INTT_barrel}
\end{figure}

INTT is a two-layer barrel silicon strip tracker. The schematic drawing of INTT is shown in Fig.~\ref{fig:INTT_barrel}. The INTT consists of 56 silicon ladders~\cite{INTT_ladder_NIM}—24 in the inner barrel and 32 in the outer barrel—arranged in a cylindrical configuration around the beam pipe, with radii of approximately 7.5\cm (inner barrel) and 10\cm (outer barrel). In each barrel, the ladders are staggered in the $\phi$ direction to ensure full azimuthal coverage. The INTT is positioned between the MAPS-based Vertex Detector (MVTX) and the Time Projection Chamber (TPC) in sPHENIX. By providing two spatial points per track, the INTT is able to bridge tracks of the MVTX and TPC to enhance pattern recognition.
In addition, the INTT plays a unique and crucial role in associating reconstructed tracks with individual RHIC bunch crossings of \qty{106}{\nano \second}, thereby enabling effective out-of-time pileup discrimination and suppression, thanks to the superior timing resolution of INTT among the \sphenix tracking detectors\footnote{The timing resolutions of the INTT, MVTX, and TPC are on the order of \qty{0.1}{\micro \second}, \qty{1}{\micro \second}, and \qty{10}{\micro \second}, respectively.}. 

\begin{figure*}[htb]
    \centering
    \includegraphics[width=1.0\linewidth]{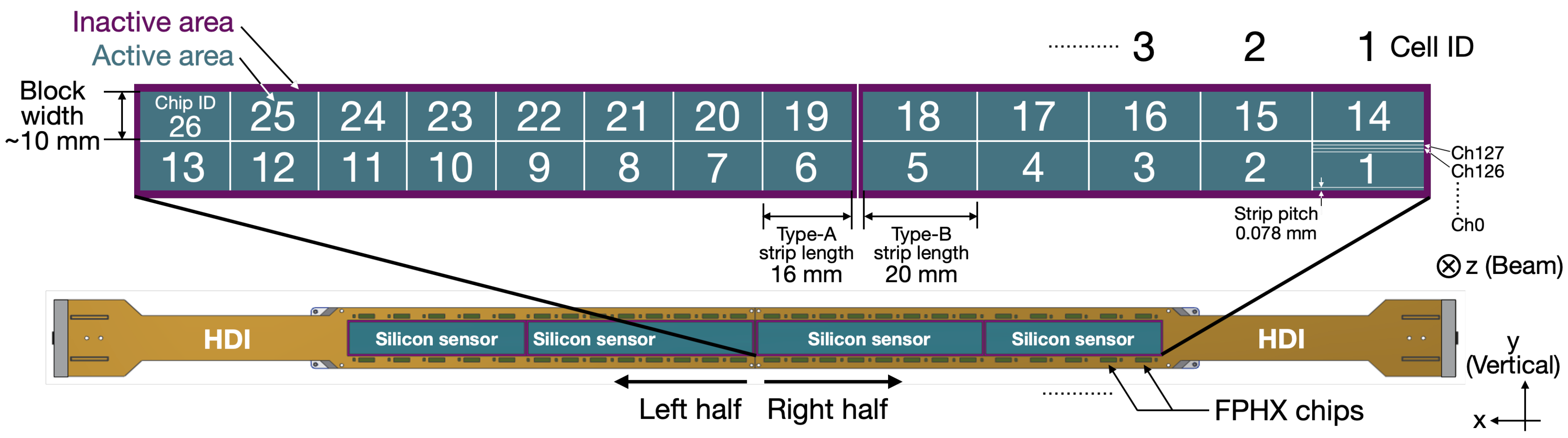}
    \caption{The schematic drawing of an INTT ladder with sensors facing up. An INTT ladder consists of two types of silicon sensors, FPHX chips, HDI cables, and a CFC stave. The sensors are divided into 16 (Type-A) and 10 (Type-B) blocks read out by individual FPHX chips. A silicon block has 128 strips oriented horizontally with a pitch of \SI{78}{\micro\meter} and a length of 16\mm (Type-A) or 20\mm (Type-B). The coordinate system used in the beam test is also shown.}
    \label{fig:INTT_ladder}
\end{figure*}

Figure~\ref{fig:INTT_ladder} presents a schematic drawing of an INTT ladder. An INTT ladder consists of Type-A and Type-B silicon sensors, 52 FPHX readout chips~\cite{PHENIX_FPHX_cite}, two high-density interconnect (HDI) cables, and a carbon fiber composite (CFC) stave underneath the HDIs as a support structure and heat transmitter. The silicon sensor is divided into sixteen (Type-A) or ten (Type-B) blocks, read out by individual FPHX chips. Each block contains 128 silicon strips with a pitch of \SI{78}{\micro\meter} and a length of 16\mm (Type-A) or 20\mm (Type-B). Details of the INTT ladder are discussed in Ref.~\cite{INTT_ladder_NIM}. Ladders are assembled at BNL and the Taiwan Instrumentation and Detector Consortium. Table~\ref{tab:INTT_spec} summarizes key specifications of the INTT ladder. 

When a charged particle traverses a silicon sensor of INTT, electron-hole pairs are created. The corresponding FPHX chip reads the analog signal and converts it to digital information. The signal amplitude is digitized using a 3-bit analog-to-digital converter (ADC), which comprises eight programmable comparators. The threshold of each comparator, hereinafter referred to as the digital-to-analog converter (DAC) value, can be set within a range of 0 to 255. The amplitude of the analog signal is compared against all comparators, and the digitized value is determined by the index of the comparator with the highest threshold exceeded. Signals with amplitudes below the lowest comparator threshold are discarded. The digitized signal is sent to the readout card (ROC) through the HDI cable and an extension cable. The left and right halves of the INTT ladder, as illustrated in Fig.~\ref{fig:INTT_ladder}, have the same structure but are operated independently.

\subsection{The beam test}

\begin{figure}[ht]
    \centering
    \includegraphics[width=\PlotsizeL\linewidth]{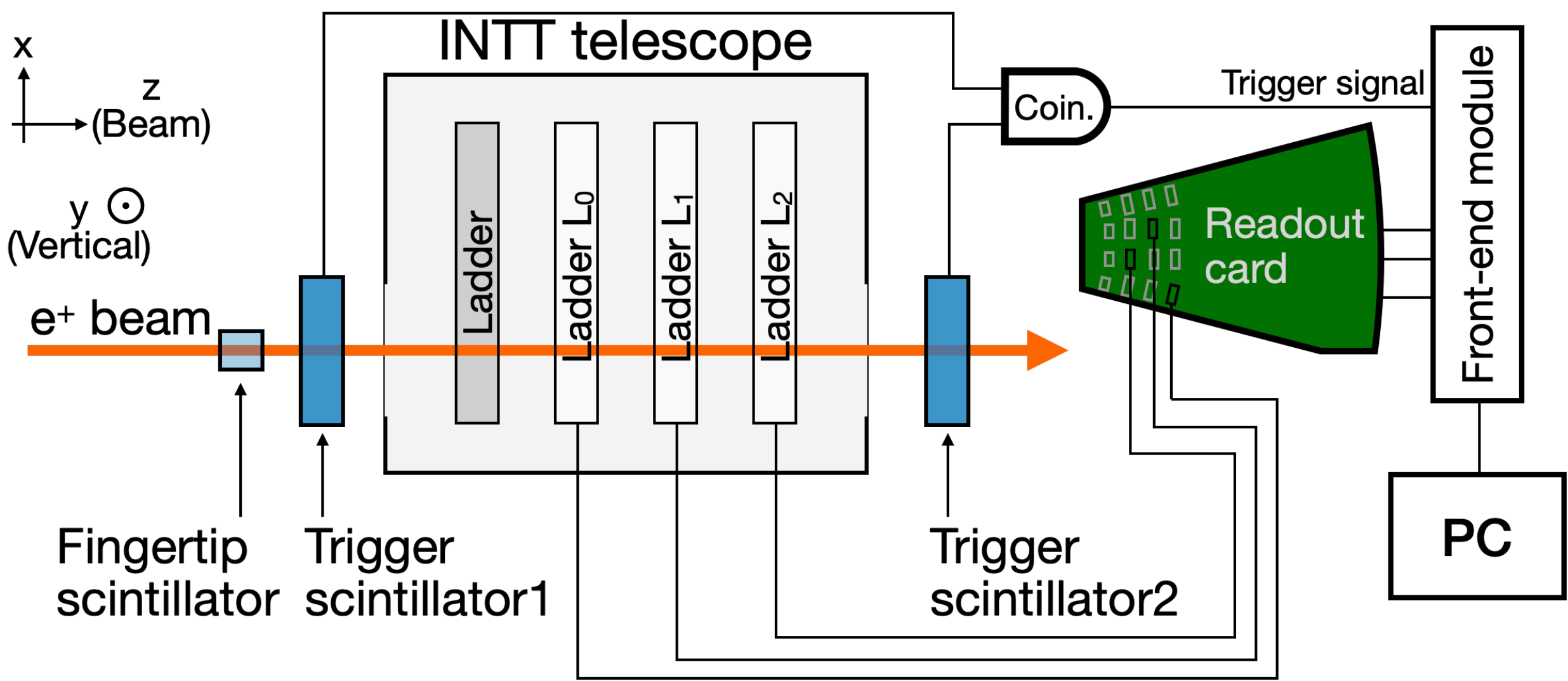}
    \caption{The schematic diagram of the setup for the INTT beam test at RARiS. The INTT telescope, containing four evenly spaced INTT preproduction ladders, was located on the positron beamline. Two trigger scintillators were installed upstream and downstream of the INTT telescope, respectively. A fingertip-sized scintillator was placed in front of the upstream trigger scintillator, but its data was not included in the analysis. The x-, y-, and z-axes used in the beam test are also shown.}
    \label{fig:set_up_drawing}
\end{figure}

A silicon tracking detector is expected to have a good signal-to-noise ratio and a high hit-detection efficiency ($>$99$\%$). A beam test was conducted to evaluate the performance of the INTT ladder and its readout chain towards the last phase of R\&D. The schematic diagram of the beam test setup is shown in Fig.~\ref{fig:set_up_drawing}. A dedicated telescope\textemdash comprising four preproduction INTT ladders evenly spaced in a dark box\textemdash was built for the beam test at the Research Center for Accelerator and Radioisotope Science\footnote{Formerly known as the research center for ELectron PHoton science, ELPH.} (RARiS)~\cite{RARiS_cite}, Tohoku University, Japan. Trigger scintillators, matching the dimensions of the silicon sensors on half of a ladder, were installed at the upstream and downstream ends of the INTT telescope. The coincidence signal of the two trigger scintillators was used as the trigger signal. A fingertip-sized scintillator was placed in front of the upstream trigger scintillator, but its data was not included in the analysis. The INTT telescope was installed on the \SI{-23}{\degree} beamline in the gamma-ray irradiation hall of RARiS. The positron beam with a momentum of 1 GeV/$c$ was produced by the interaction of the primary gamma-ray beam with a \SI{200}{\micro\meter}-thick tungsten production target.

During the beam test, the right halves of the INTT ladders were operated for data collection, except for the most upstream ladder, due to a bias voltage issue\footnote{Though the most upstream ladder was not operated during the beam test, its water-cooling tube was connected in series with other ladders. It was therefore left in place on the beamline without being removed.}. The remaining three ladders showed good performance and are denoted as \ladder{0}, \ladder{1}, and \ladder{2} hereafter. In each FPHX chip, the lowest comparator threshold was set to a DAC value of 15, unless otherwise specified. The trigger signals (processed by NIM logic modules) and the INTT hit data were sent to a PHENIX FVTX front-end module (FEM)~\cite{AIDALA201444}. The FEM stores the INTT data only when an active trigger signal is present. The data were collected using a data acquisition system developed with the PCIe-6536B (National Instruments Co.) running on a Windows 10 operating system. 

We note that the INTT operational configuration used in the beam test was a prototype setup and differed slightly from the final configuration for the sPHENIX detector. In the beam test, a 40-centimeter flexible printed circuit (FPC)-based cable was used to transmit signals between the HDI and ROC during the measurements, except for one run used for the hit-detection efficiency study (as discussed in Section~\ref{sec_detection_effi}). In this run, a 1.3-meter prototype bus-extender (BEX) cable~\cite{cite_BEX_IEEE} was additionally connected to \ladder{1} following the FPC-based cable. This measurement was intended to assess the reliability of data transmission through the BEX. Data readout was performed using an FEM.  In the final INTT setup, a \textmu-coaxial cable~\cite{INTT_ladder_NIM} and a shorter 1.11-meter bus-extender cable are employed for each HDI-to-ROC connection, with data read out by the FELIX board~\cite{Anderson_2016}. We also note that the bias voltage applied to the sensors during the beam test was 50 V, slightly below the full depletion voltage of approximately 57 V. This was not an optimal configuration. Nevertheless, as will be discussed in Section~\ref{Analysis_and_results}, the impact on the performance presented in this paper (hit-detection efficiency, multiple track reconstruction, etc.) under a bias voltage slightly below the full depletion voltage is marginal. For INTT during the sPHENIX data taking, a nominal operating bias voltage of 100 V is applied.

\begin{table}[ht]
    \centering
    \caption{Specification of the INTT ladder.}
    \begin{tabular}{c c}
        \toprule[1pt]
        \textbf{Element} & \textbf{Value} \\
        \midrule[1pt]
        Radiation length & 1.14\% [\ensuremath{X/X_0}] \\
        Active area & $45.6 \times 2 \cma$  \\
        Number of channels & 6,656 \\
        Channel strip pitch & \SI{78}{\micro\meter} \\
        \multirow{2}{*}{Channel strip length} & 16\mm (Type-A sensor) \\
         & 20\mm (Type-B sensor) \\
        \bottomrule[1pt]
    \end{tabular}
    \label{tab:INTT_spec}
\end{table}
\section{Analysis and results}
\label{Analysis_and_results}

The section is organized into four topics: the signal-to-noise ratio (Section~\ref{sec_edep}), the residual distribution and the INTT spatial resolution (Section~\ref{sec_residual}), the hit-detection efficiency (Section~\ref{sec_detection_effi}), and the multiple track reconstruction (Section~\ref{sec_NTrack}). Aspects of the analysis common to the topics are introduced jointly (Section~\ref{sec_common_ana}), while topic-specific methods and results are discussed in the respective subsections.

\subsection{Common analysis components}
\label{sec_common_ana}

Three noisy channels found in \ladder{2} are excluded from the analysis\footnote{The noisy channels are far away from the beam-spot.}. Vertically adjacent blocks in each ladder, for example, chips 1 and 14 in Fig.~\ref{fig:INTT_ladder}, are treated as one cell. Amplitudes of the recorded INTT hits are converted to the predefined threshold setting of the corresponding comparator, after which clustering is performed. In a given cell of a ladder, groups of vertically adjacent hits are treated as clusters. The cluster's x-position is assigned as the same cell index, while its y-position, $Y$, is determined using the weighted average method:
\begin{equation*}
    \begin{aligned}
        &Y = \sum\limits_{i=1}^{n}(e_{i} \cdot y_{i}) / \sum\limits_{i=1}^{n}e_{i},
    \end{aligned}
    \label{eq:energy_weight}
\end{equation*}
where $n$ is the number of hits in the cluster, and $e_{i}$ and $y_{i}$ denote the converted amplitude and y-position of the $i$-th hit. The cluster amplitude is defined as the sum of the $e_i$ values of the hits in the cluster. These clusters represent the total energy deposited by a charged particle traversing an INTT ladder and carry information about its location. The resulting clusters serve as input to subsequent analyses.

To compare the INTT ladder performance with simulation, a \geant~\cite{Geant4_cite} model is developed based on the INTT beam test setup, including the INTT telescope\footnote{The most upstream ladder that was not operated during the beam test was also included in the model.} and the scintillator configuration. In addition, the particle-gun generator is configured according to the beam characteristics of the \SI{-23}{\degree} beamline in the gamma-ray irradiation hall of RARiS.

The tracking analysis begins in Section~\ref{sec_residual}. It includes correlating clusters across multiple ladders in the selected cell and correcting the ladder misalignment along the vertical axis. Given the setup of the INTT telescope, a residual can be defined as $r = (Y_{\mathrm{L_{1}}}+C_{\mathrm{L_{1}}}) - Y_{\mathrm{pred}}$, where $Y_{\mathrm{L_{1}}}$ is the position of the \ladder{1} cluster in the y-axis, $C_{\mathrm{L_{1}}}$ is the misalignment correction for \ladder{1} relative to \ladder{0} and \ladder{2}, and $Y_{\mathrm{pred}}$ is the expected y-position for the \ladder{1} cluster, obtained by a linear interpolation of \ladder{0} and \ladder{2} clusters. 

\begin{figure}
    \centering
    \includegraphics[width=\Plotsize\linewidth]{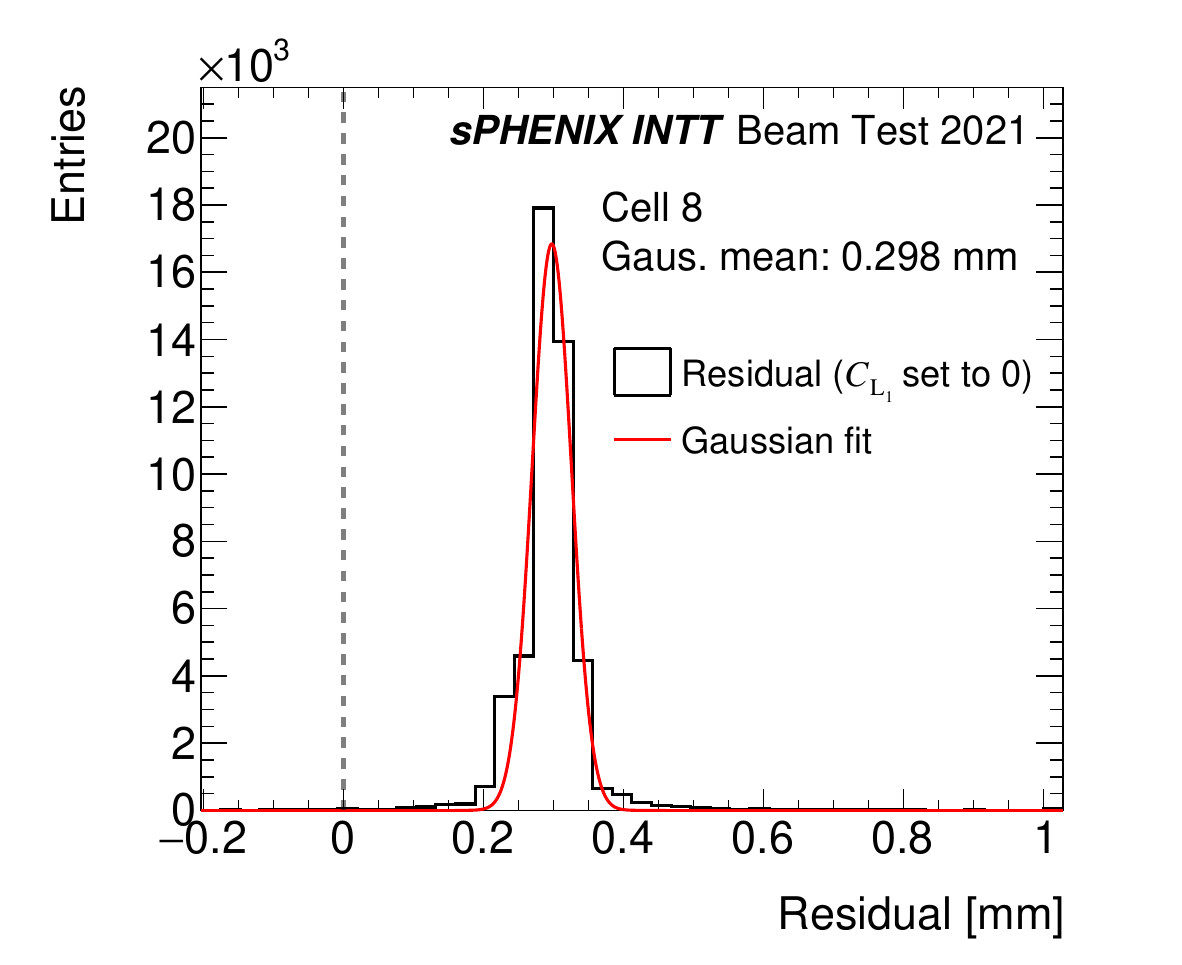}
    \caption{The distribution filled by $r$ with $C_{\mathrm{L_{1}}}$ set to zero for a representative cell. The mean of the Gaussian function fitted to the distribution indicates the amount of misalignment of \ladder{1} relative to \ladder{0} and \ladder{2}.}
    \label{fig:residual_alignment}
\end{figure}

 The $C_{\mathrm{L_{1}}}$ is determined using events with exactly one cluster in the selected cell of each ladder. In addition, the absence of clusters in cells adjacent to the selected one is required to account for potential misalignment along the horizontal axis. The $r$ is then calculated with $C_{\mathrm{L_{1}}}$ set to zero. The resulting distribution of a representative cell is shown in Fig.~\ref{fig:residual_alignment}. The mean of a Gaussian function fitted to the distribution is offset from zero by 0.298 mm, indicating that \ladder{1} is positioned slightly lower relative to \ladder{0} and \ladder{2} in the vertical axis. Accordingly, $C_{\mathrm{L_{1}}}$ is assigned to be –0.298 mm, the negative Gaussian mean, to correct for the misalignment of \ladder{1}.

\subsection{Signal-to-noise ratio}
\label{sec_edep}

The signal-to-noise ratio is one of the key parameters for evaluating the performance of a tracking detector and ensuring the reliability of the recorded data. The signal-to-noise ratio is $R \equiv S/N$, where $S$ is the most probable energy deposition of minimum ionizing particles, and $N$ is the root mean square of the noise distribution of the system. The $R$  can be extracted from an energy-deposit distribution. As a trade-off for the low power consumption, the energy resolution of the FPHX chip is limited by its 3-bit ADC, which is insufficient to study the energy-deposit distribution with high precision. To address this, a series of eight narrow-range measurements, with threshold increments as small as 4 DAC, referred to as a DAC scan, was conducted. The measurement ranges of the DAC scan runs are summarized in Table~\ref{tab:DAC_scan_range}. This approach allows each measurement to cover a slightly different region of the energy-deposit distribution, with some overlap between adjacent ones, enabling the full energy-deposit spectrum to be mapped with high precision.

\begin{figure}[ht]
    \centering
    \includegraphics[width=\Plotsize\linewidth]{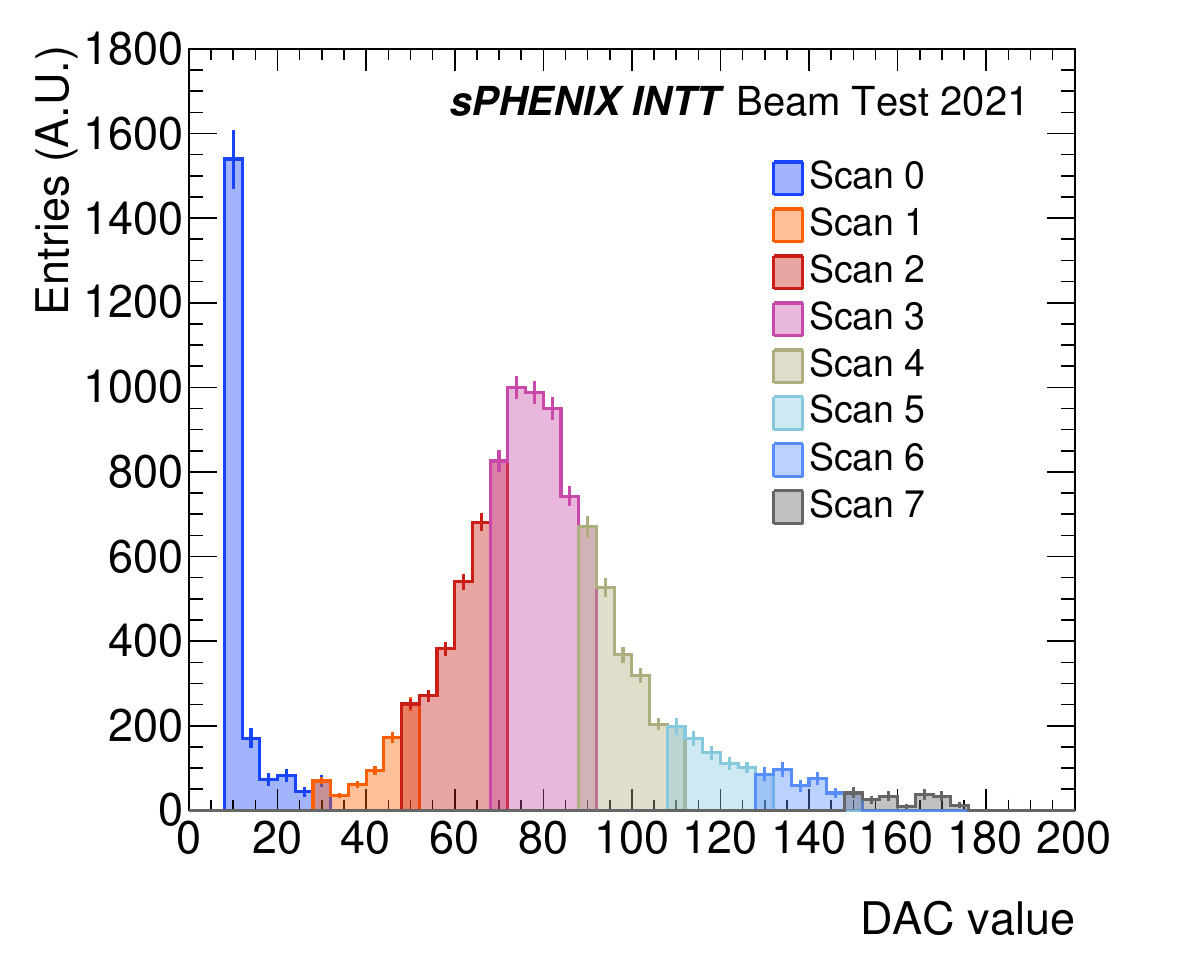}
    \caption{The energy-deposit distribution, constructed by the eight sequential measurements (histograms with different colors) after applying scaling factors, for a representative ladder (\ladder{0}).}
    \label{fig:l0_edep_overlap}
\end{figure}

Ladders \ladder{0} and \ladder{2} are included in this study. Events with at most one cluster per ladder are selected. The analysis focuses on clusters in the beam-spot region of the selected events. Amplitudes of single-hit clusters (clusters consisting of a single activated channel, with no adjacent channels registering a hit) are binned into separate histograms for each measurement. Scaling factors derived from overlapping bins are applied to the histograms to normalize the eight measurements and construct the full energy-deposit distribution, as illustrated in Fig.~\ref{fig:l0_edep_overlap}. 

\begin{figure}[ht]
    \centering
    \includegraphics[width=\Plotsize\linewidth]{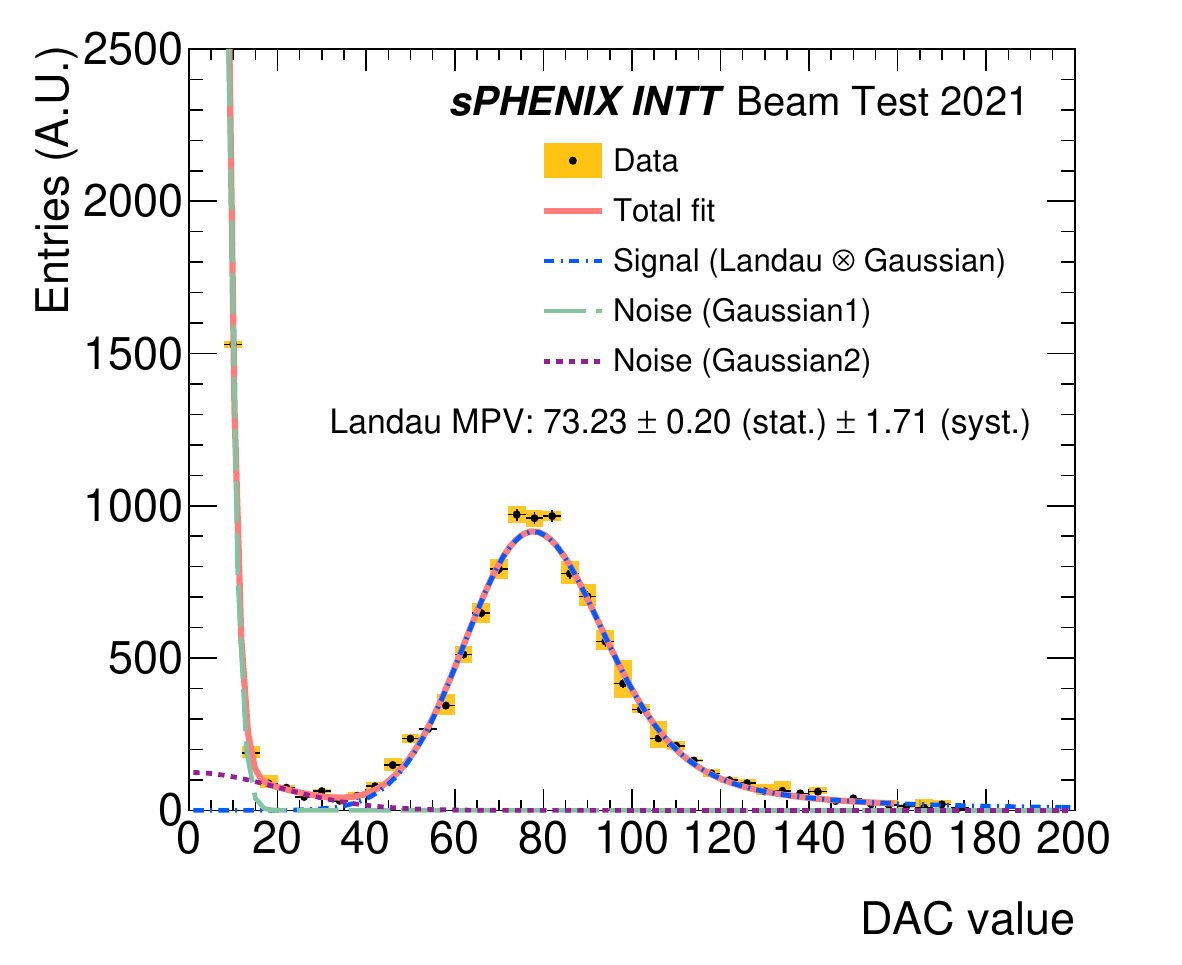}
    \caption{The energy-deposit distribution of the INTT ladders fitted by a Landau-Gaussian convolution function for the signal component, plus two Gaussian functions for the noise component. The vertical extent of each yellow box represents the systematic uncertainty of the ladder variation, while the vertical bar shows the statistical uncertainty.}
    \label{fig:edep_final}
\end{figure}

For each of \ladder{0} and \ladder{2}, the eight corresponding distributions are merged to produce an integrated distribution, referred to hereafter as $E_{\mathrm{0}}$ and $E_{\mathrm{2}}$. They are then statistically combined to obtain the average distribution, representing the average energy-deposit distribution of the INTT ladders, as shown in Fig.~\ref{fig:edep_final}. Deviations of $E_{\mathrm{0}}$ and $E_{\mathrm{2}}$ from the average are assigned as systematic uncertainties (yellow boxes). A clear signal component followed by a steeply falling noise component is observed. The average distribution, as well as $E_{\mathrm{0}}$ and $E_{\mathrm{2}}$, is fitted with a Landau-Gaussian convolution function (Landau $\otimes$ Gaussian)~\cite{PhysRevA.28.615} for the signal component, plus two Gaussian functions with peaks fixed at the origin for the noise component. Taking the variation of the Landau most probable values (MPV) of $E_{\mathrm{0}}$ and $E_{\mathrm{2}}$ as a source of systematic uncertainty, the most probable energy deposition of minimum ionizing particles for the INTT ladders is measured to be a DAC value of $\mathrm{73.23 \pm 0.20 \; (stat.) \pm 1.71 \;(syst.)}$. The noise width is determined as the average of the widths of the two Gaussian functions fitted to the average distribution, weighted by their respective areas. The resulting noise width corresponds to a DAC value of $4.56\pm0.16$. The signal-to-noise ratio of the INTT ladders is measured to be greater than 15.1.

\begin{table}[ht]
    \centering
    \caption{The measurement ranges of DAC scan runs used for constructing the INTT energy-deposit distribution.}
    \begin{tabular}{c c c}
        \toprule[1pt]
        \textbf{Scan} & \textbf{Minimum} & \textbf{Maximum} \\
        \midrule[1pt]
        0 & 8 & 36 \\
        1 & 28 & 56 \\
        2 & 48 & 76 \\
        3 & 68 & 96 \\
        4 & 88 & 116 \\
        5 & 108 & 136 \\
        6 & 128 & 156 \\
        7 & 148 & 176 \\
        \bottomrule[1pt]
    \end{tabular}
    \label{tab:DAC_scan_range}
\end{table}
\subsection{Residual distribution}
\label{sec_residual}

\begin{figure}
    \centering
    \includegraphics[width=\Plotsize\linewidth]{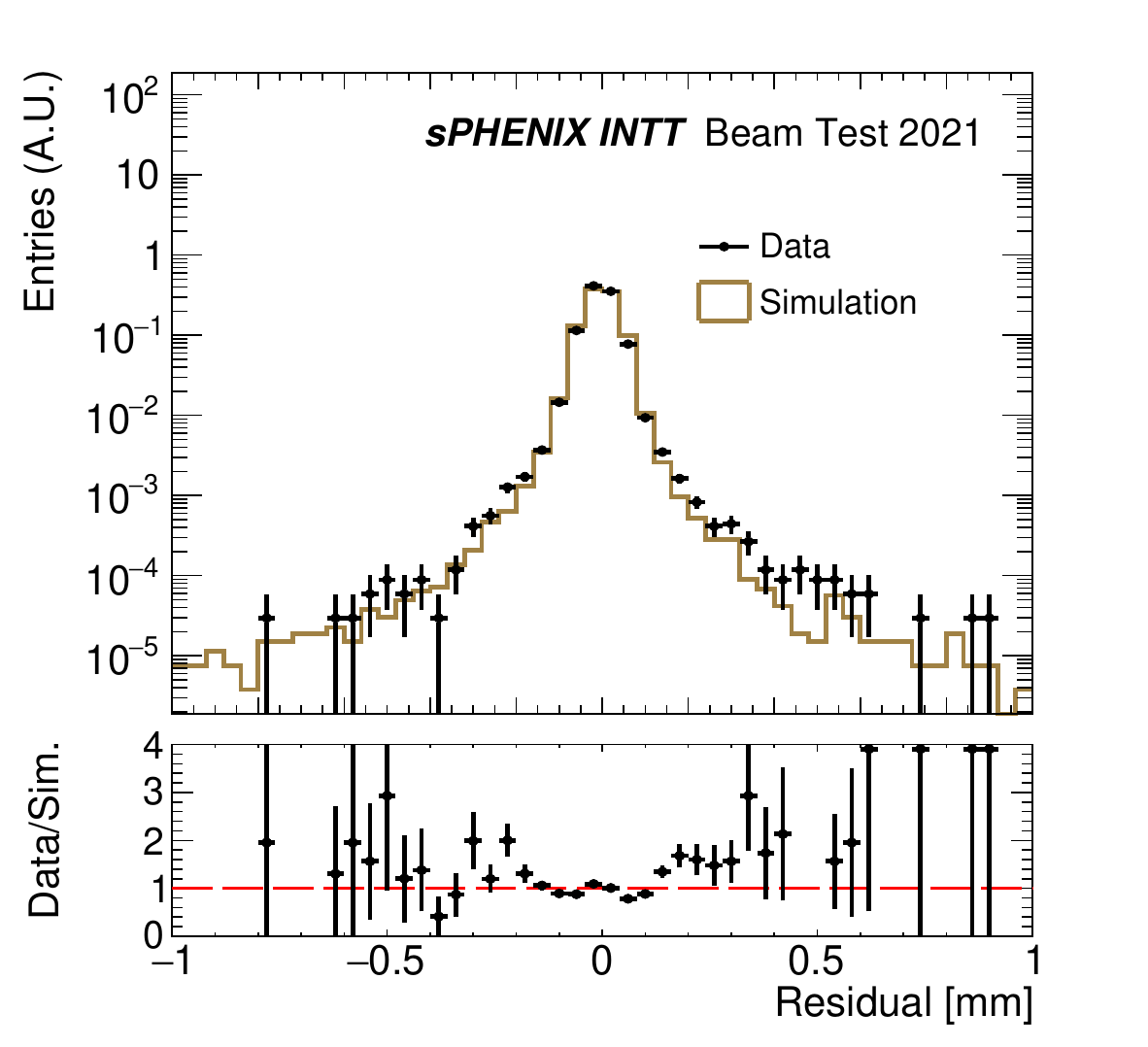}
    \caption{Residual distributions obtained from data (solid black circles) and simulation (brown line). The bottom panel presents the ratio of the data to the simulation.}
    \label{fig:residual_final}
\end{figure}

To obtain the residual distribution, a set of event selection criteria is applied. Events with one cluster per ladder, located in the selected cell, are retained. Additionally, to ensure the clusters originate from beam particles, the \ladder{0} and \ladder{1} clusters must lie within the beam-spot, and the vertical slope of the line connecting them is required to be close to zero, matching that of the beam direction.  Residuals are calculated as the outcome of tracking, in which clusters in three ladders are correlated. The results are shown in Fig.~\ref{fig:residual_final}. The data exhibit a symmetric distribution with steeply falling tails, owing to the low material budget of the INTT ladders. The \ladder{1} misalignment measured from the data was incorporated into the \geant model. A distribution obtained from the simulation through the same analysis procedure was prepared for comparison. The residual distributions in data and simulation are in agreement, which indicates good control of the implemented geometry and material configuration.

\subsubsection{Spatial resolution}
We note that it is not straightforward to extract the spatial resolution of the INTT ladder along the y-axis directly from the residual distribution of a full \geant simulation, because the residual distribution includes contributions from the intrinsic resolution of the INTT ladder ($\sigma_{\mathrm{lad}}$) and the effect of multiple Coulomb scattering. Therefore, an analytical approach is employed, as described below. 

The resolution of $Y_{\mathrm{pred}}$ ($\sigma_{Y_{\mathrm{pred}}}$) is given as: 
\begin{equation*}
    \label{resolution_eq_2}
    \begin{aligned}
    \sigma_{Y_{\mathrm{pred}}}^{2} &= \frac{1}{2}\cdot\sigma_{\mathrm{lad}}^{2}\mathrm{,}
    \end{aligned}
\end{equation*} 
following the formula outlined in Ref.~\cite{Resolution_cite}. The measured residual distribution effectively corresponds to the residual distribution of the multiple Coulomb scattering (MS) convolved with a Gaussian distribution of width $\sqrt{3/2} \cdot \sigma_{\mathrm{lad}}$. This Gaussian distribution reflects the finite spatial resolution of the three INTT ladders in the INTT telescope.

\begin{figure}
    \centering
    \includegraphics[width=\Plotsize\linewidth]{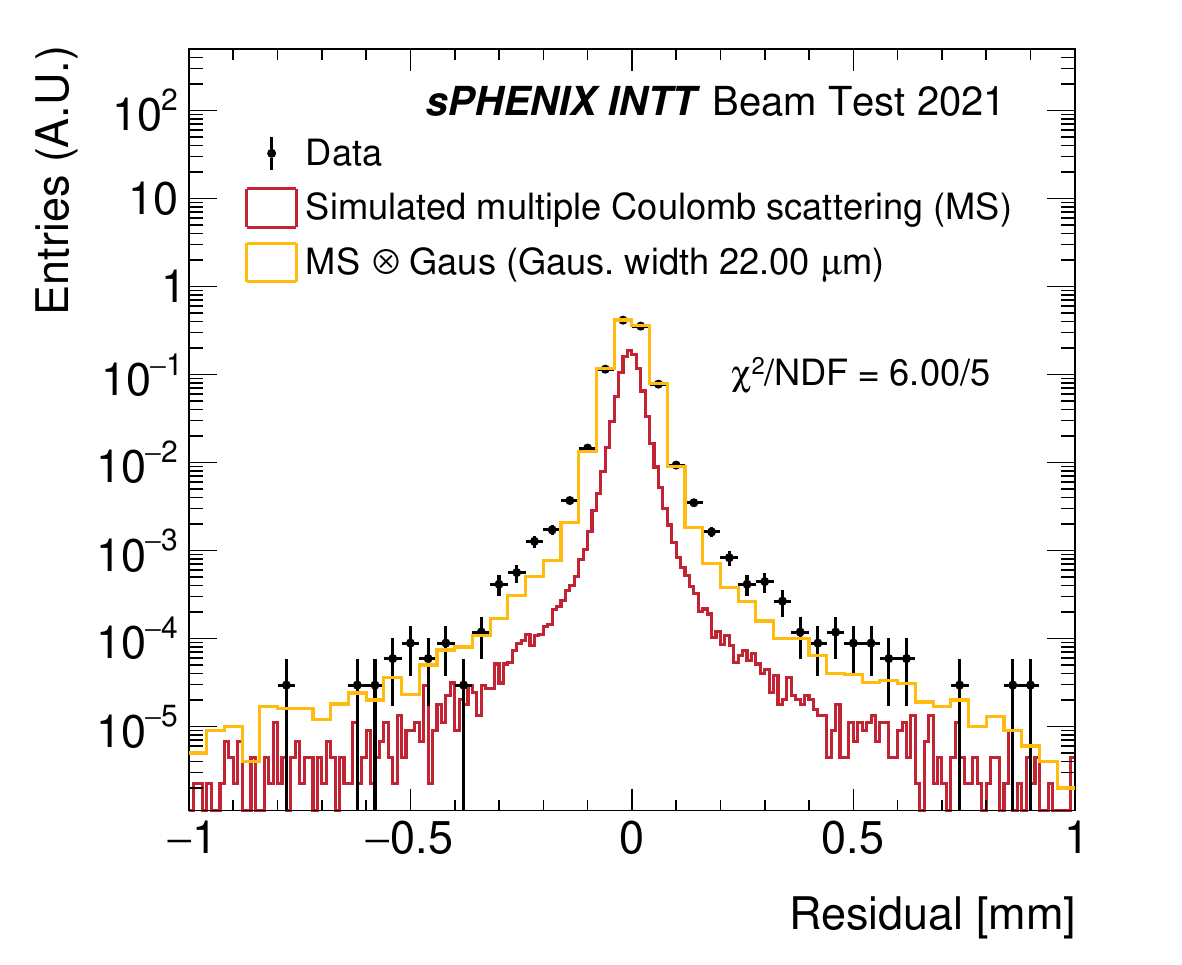}
    \caption{A representative example of a residual distribution (yellow line), obtained by convolving a simulated residual distribution of multiple Coulomb scattering (red line) with a Gaussian distribution of width \SI{22}{\micro\meter}, compared with the data distribution (solid black circles). The agreement is quantified by a chi-square test, using the central six bins that contain over 98\% of the entries.}
    \label{fig:spatial_scan_example}
\end{figure}

\begin{figure}
    \centering
    \includegraphics[width=\Plotsize\linewidth]{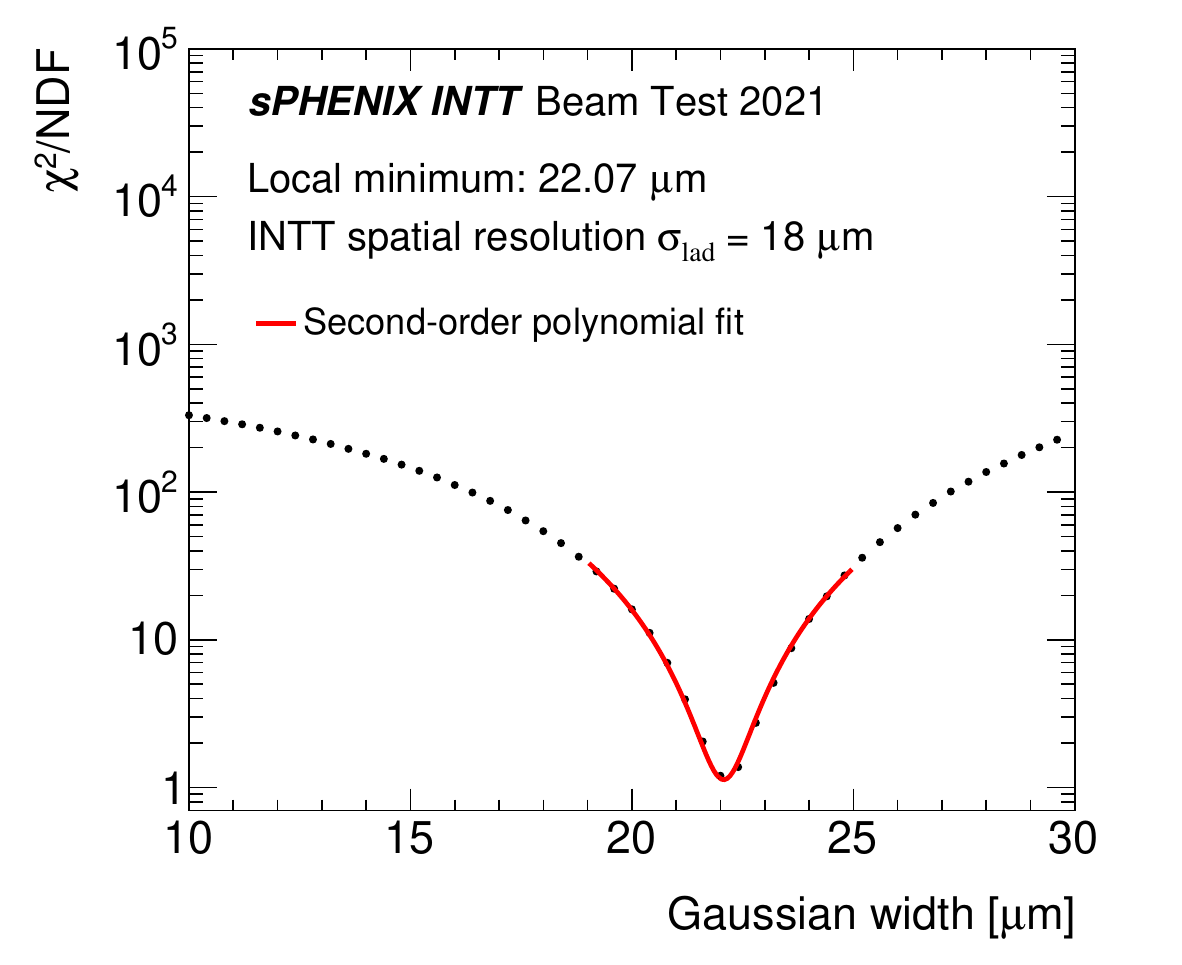}
    \caption{The reduced $\chi^{2}$ as a function of the Gaussian width. A second-order polynomial function (red line) is fitted to the curve to obtain the local minimum at \SI{22.07}{\micro\meter}, corresponding to an INTT spatial resolution of \SI{18}{\micro\meter}.}
    \label{fig:spatial_final}
\end{figure}

A scan is performed by convolving the simulated residual distribution of the multiple Coulomb scattering with a series of Gaussian functions of increasing width (MS $\otimes$ Gaus), and the resulting distributions are compared with the data. A representative example is given in Fig.~\ref{fig:spatial_scan_example}. The chi-square test is employed to quantify the agreement, using the central six bins that contain over 98\% of the entries. The reduced $\chi^{2}$ as a function of the Gaussian width is shown in Fig.~\ref{fig:spatial_final}. A second-order polynomial function (red line) is fitted to the curve. A local minimum at \SI{22.07}{\micro\meter} is observed, corresponding to an INTT spatial resolution, $\sigma_{\mathrm{lad}}$, of \SI{18}{\micro\meter}. 

\subsection{Hit-detection efficiency}
\label{sec_detection_effi}

The hit-detection efficiency of silicon tracking detectors, denoted as $\epsilon$, is generally expected to be high (>99\%) due to their favorable signal-to-noise ratio. Typically, $\epsilon$ of a device under test (DUT) is evaluated using a tracking system. The DUT is positioned between the upstream and downstream sectors of the tracking system. When a good track is reconstructed, the presence or absence of a corresponding hit in the DUT allows for an assessment of its detection efficiency, as the track is known to have traversed the DUT. In this beam test, \ladder{0} and \ladder{2} are used for track reconstruction, while \ladder{1} is designated as the DUT, and its hit-detection efficiency is measured.

Since tracks are reconstructed using only two ladders (\ladder{0} and \ladder{2}), stringent event selection criteria are applied to ensure the purity of reconstructed tracks. The cell on the beam-spot is chosen as the baseline. To minimize track ambiguity and account for potential ladder misalignment along the horizontal axis, events are required to have exactly one cluster in the selected cell of \ladder{0} and \ladder{2}, and no clusters in the adjacent cells of any of the three ladders. In addition, events in which the smallest-amplitude cluster occurs in \ladder{0} or \ladder{2} are discarded to suppress noise contribution. Furthermore, the vertical slope of the line connecting the clusters of \ladder{0} and \ladder{2} is required to be close to zero to align with the beam direction, and the y-positions of these clusters must be within the beam-spot region to ensure the reconstructed tracks originate from the beam.

Events that pass these criteria are considered to contain a good track. $Y_{\mathrm{pred}}$ is then obtained from the track, and clusters in the selected cell of \ladder{1} are examined. If a cluster is found in \ladder{1} and its absolute residual is within 0.7\mm, the event is classified as a successful detection of a particle hit in \ladder{1}. The residual criterion is determined from a simulation study, in which the detection efficiency exceeds $99.995\%$ for $|r| < \mathrm{0.7}$\mm. The hit-detection efficiency is defined as: 
\begin{equation*}
    \begin{aligned}
     \epsilon &= \frac{\text{Events with $\mathrm{L_{1}}$ cluster matching to track}}{\text{Events with good track}}.
    \end{aligned}
    \label{eq:effi_equation}
\end{equation*} 

 Systematic uncertainties affecting the efficiency measurement are evaluated. The stability of detector components is assessed by performing the analysis with the cell having the second-highest cluster count. The sensitivity of the selection for the track's vertical slope is examined by varying the criterion in the selection. For each source, the resulting deviation in hit-detection efficiency is quoted as a systematic uncertainty.
 
\begin{figure}
     \centering
     \includegraphics[width=\Plotsize\linewidth]{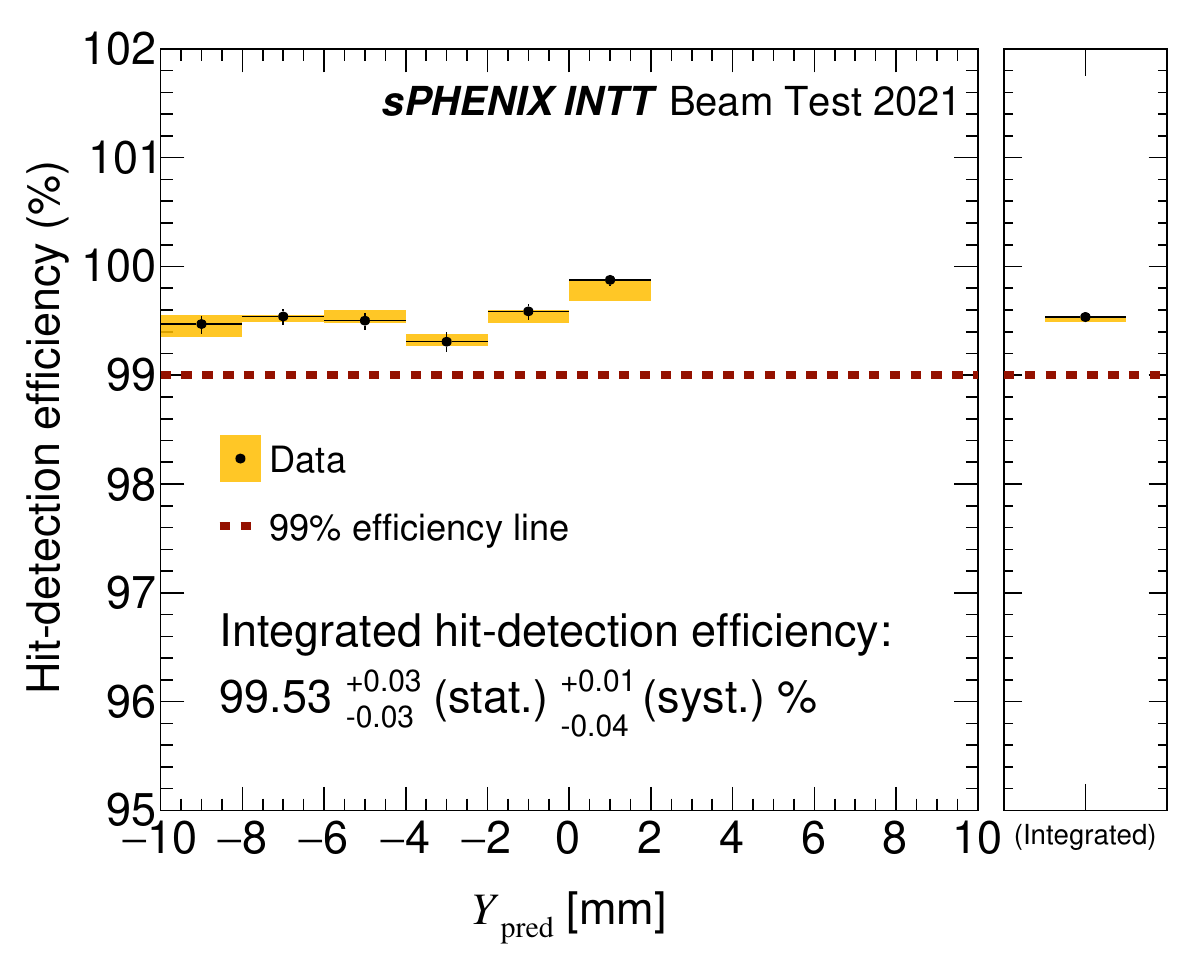}
     \caption{The hit-detection efficiency of \ladder{1} as a function of $Y_{\mathrm{pred}}$ (left panel), as well as the integrated result (right panel). The vertical extent of each yellow box represents the total systematic uncertainty, while the vertical bar shows the statistical uncertainty. The red dashed line represents an efficiency of 99$\%$.}
     \label{fig:DetectionEffi_Final}
 \end{figure}

 Figure~\ref{fig:DetectionEffi_Final} presents the hit-detection efficiency of \ladder{1} as a function of $Y_{\mathrm{pred}}$ (left panel), as well as the integrated result (right panel). The total systematic uncertainty (yellow boxes) is obtained by summing all sources in quadrature under the assumption that they are independent and uncorrelated. The data cover the y-axis range from \SI{-10}{mm} to +\SI{2}{mm}, partially spanning the sensor's full extent (\SI{-10}{mm} to +\SI{10}{mm}). While the entire sensor is not covered, the data reach the edge on one side. Within the reported range, $\epsilon$ stays above 99$\%$. The integrated hit-detection efficiency of \ladder{1} is measured to be  $\mathrm{99.53\;^{+0.03}_{-0.03} \; (stat.)\;^{+0.01}_{-0.04} \; (syst.) \%}$. 
 
 The analysis is also performed using another run with the nominal cabling configuration as a reference. The hit-detection efficiencies obtained from the two runs differ by less than 0.1\%, indicating excellent reliability of the data transmission through the BEX.

\subsection{Multiple track reconstruction}
\label{sec_NTrack}

The performance of multiple track reconstruction is demonstrated using a special run with an additional 1~cm-thick lead plate (1.78 \ensuremath{X/X_0}) placed 40~cm upstream of the INTT telescope. The positron beam is delivered and directed onto the lead plate, inducing a particle shower. Secondary particles from this interaction are expected to traverse the INTT telescope. All 13 cells are included in this analysis to maximize the detector acceptance. The track-finding procedure starts by looping over all clusters in a given cell across the three INTT ladders. The most linear combination, quantified by a straight-line fit, is selected. If the residual $r$ of the chosen combination is within $\pm$0.7\mm, it is classified as a good track. The associated clusters are removed from the rest of the track-finding for a given event, and the procedure is repeated iteratively until no further combination satisfies the residual criterion.

\begin{figure} 
    \centering
    \includegraphics[width=\Plotsize\linewidth]{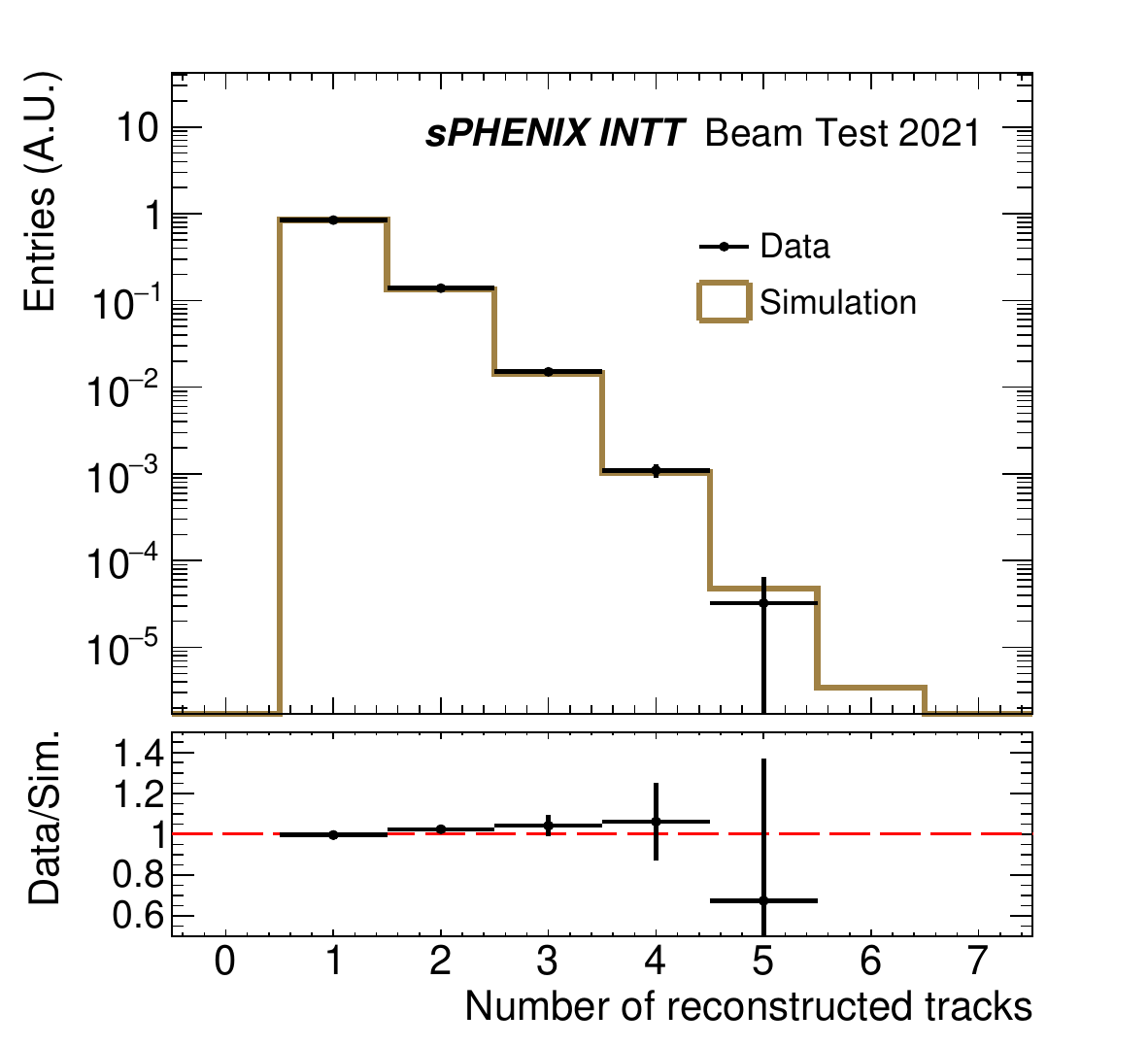}
    \caption{Distributions of the number of reconstructed tracks using the full acceptance available for the beam test in data (solid black circles) and simulation (brown line). The bottom panel presents the ratio of the data to the simulation.}
    \label{fig:NTrack_final}
\end{figure}

Figure~\ref{fig:NTrack_final} shows the distribution of the number of reconstructed tracks per event. Given the full acceptance available for the beam test, up to five tracks can be reconstructed from a single event. The same distribution from a dedicated simulation sample using the same analysis procedure is presented for comparison. The data are in good agreement with the simulation, serving as a benchmark measurement that validates the tracking performance of the INTT under high-multiplicity conditions.

\section{Conclusion}

The performance of the Intermediate Silicon Tracker (INTT) was evaluated using a positron beam with a momentum of 1 GeV/$c$ at RARiS. The INTT ladders demonstrated a signal-to-noise ratio exceeding 15.1, and a hit-detection efficiency above 99$\%$ in the reported range along the vertical axis. The measured residual distribution showed strong agreement with the simulation, indicating good control of the ladder geometry and material configuration, as well as high data quality. The spatial resolution of the INTT, evaluated from the residual distribution, was measured to be \SI{18}{\micro\meter}. Multiple track reconstruction performance was further benchmarked using a positron-induced particle shower, with data and simulation in good agreement. These results confirm that the INTT meets the performance requirements for precision tracking. The INTT was installed and commissioned in the sPHENIX detector in 2023 and has since been collecting data stably with p+p and Au+Au collisions~\cite{shih2025intermediate}, leading to the first physics results from sPHENIX~\cite{sPHENIX:2025uhi}.


\section*{Acknowledgments}

We express our gratitude to RARiS and the staff for the stable operation of the accelerators during the beam test. We are grateful to the sPHENIX collaboration for their support and assistance during various phases of the INTT ladder and readout cable developments. We thank the PHENIX FVTX experts for providing essential documentation and devices. We thank Prof. Takatsugu Ishikawa for serving as the liaison between RARiS and our group during the beam test and for providing the simulation setup used to model the beam characteristics of the \SI{-23}{\degree} beamline of the gamma-ray irradiation hall of RARiS.

We acknowledge support from the Ministry of Education, Culture, Sports, Science, and Technology and the Japan Society for the Promotion of Science (Japan); Office of Nuclear Physics in the Office of Science of the Department of Energy (U.S.A.); National Science and Technology Council and the Ministry of Education (Taiwan). 

\bibliographystyle{unsrturl}
\bibliography{reference}

\end{document}